\documentstyle[11pt,paspconf,psfig]{article}

\newcommand{\etal}{{\it et al.\ }}
\newcommand{\lta}{\stackrel{<}{\scriptstyle\sim}}
\newcommand{\gta}{\stackrel{>}{\scriptstyle\sim}}
\newcommand{\R}{${\cal R}$} 

\begin{document}

\title{Chemically Consistent Evolutionary Synthesis Models}

\author{Uta Fritze - v. Alvensleben}
\affil{Universit\"atssternwarte G\"ottingen}

\begin{abstract}
Any stellar system with a star formation history (SFH) more extended 
than a massive star's lifetime will be composite in metallicity. 
Our method of chemically consistent evolutionary synthesis tries 
to account for the increasing initial metallicity of successive generations 
of stars. Using various sets of input physics for a range 
of metallicities $10^{-4} \leq Z \leq 0.05$ we keep track of the ISM 
enrichment and follow successive generations of stars using stellar 
evolutionary tracks, yields, model atmosphere spectra, index calibrations, 
etc., appropriate for their respective initial metallicities. Since the SFH 
determines the evolution not only of the metallicity, but, in particular, of 
abundance ratios of specific 
elements, stellar evolution and galaxy evolution become intimately coupled. 
I review the concept of chemically consistent evolution, present results for 
the photometric, spectral, and chemical evolution of galaxies of various types in the 
local Universe and at high redshift, and discuss its advantages as well as its current limitations. 

\end{abstract}


\keywords{galaxies, chemical evolution, spectrophotometric evolution}

\section{Introduction}
In principle, any stellar system with star formation (SF) going on over more than the lifetime 
of the most massive stars plus the cooling time of the gas is composite in terms of age and 
metallicity. Since this is the case for any composite stellar system like a galaxy -- as 
opposed to star clusters -- the importance to account for realistic metallicity distributions 
in evolutionary galaxy models is evident. For quite some time, observational evidence has been a
ccumulating for finite and often very large metallicity distributions, e.g. spanning a range of 
more than a factor of 100 in bulges from ${\rm (0.01 - 3)\cdot Z_{\odot}}$ (e.g. Jacoby \& 
Ciardullo 1999). While some years ago, the focus was on super-solar metallicities e.g. in 
(the centers of) massive ellipticals, bulges, X-halos around ellipticals, and the hot ICM, 
by today, it is clear that the {\bf average metallicities} in all those cases are {\bf subsolar}. 
When averaged over ${\rm 1~R_e}$, line strength gradients in E/S0s show 
${\rm \langle Z_{\ast} \rangle \sim (0.5 - 1)\cdot Z_{\odot}}$ (Carollo \& Danziger 1994). For 
stars in bulges ${\rm \langle Z_{\ast} \rangle \sim (0.3 - 0.7)\cdot Z_{\odot}}$ (e.g. 
McWilliam \& Rich 1994), for the X-gas halos of ellipticals ASCA observations give 
${\rm 0.1 \leq [Fe/H] \leq 0.7}$ (e.g. Loewenstein 1999). Characteristic
HII region abundances (i.e. measured at ${\rm 1~R_e}$) range from ${\rm Z \gta Z_{\odot}}$ 
for Sa spirals down to ${\rm \sim \frac{1}{2}~ Z_{\odot}}$ 
for Sd galaxies (e.g. Oey \& Kennicutt 1993, Zaritsky \etal 1994, Ferguson \etal 1998, van 
Zee \etal 1998). The sun, our reference star, stands out in metallicity among solar 
neighborhood stars. For F, G, K dwarfs the [Fe/H] distributions extend from ${\rm -0.8~ 
to~ +0.4}$ (Rocha-Pinto \& Maciel 1998), while B-stars show ${\rm \langle [O/H] \rangle = -0.31}$ 
(Kilian-Montenbruck \etal 1994). Locally already, dwarf irregulars have metallicities in 
the range (2 -- 30)\% ${\rm Z_{\odot}}$ (e.g. Richer \& McCall 1995). The first spectra of 
Lyman break galaxies at ${\rm z \sim 3 - 4}$ have shown that their metallicities, derived 
from stellar wind features, are considerably subsolar, sometimes even sub-SMC (Lowenthal 
\etal 1997, Trager \etal 1997). Neutral gas in damped Ly$\alpha$ absorbers observed to 
${\rm z > 4}$ shows abundances ${\rm -3 \lta [Zn/H] \lta 0}$ (cf. Sect.5). 

\section{Chemically Consistent Modelling}
Evolutionary synthesis models start from a gas cloud of mass G, initially comprising the total mass M, 
with primordial abundances, 
give a star formation rate ${\rm \Psi(t)}$ and an IMF, and 
form the 1$^{st}$ generation of stars with ${\rm Z = 0}$. We solve a modified form of 
Tinsley's equations with stellar yields for SNII, SNI, PN, and stellar mass loss 
for ${\rm Z = 0}$ to obtain 
ISM abundances and abundance ratios. The next generation of stars is formed 
with abundances $Z>0$ and abundance ratios [X$_i$/X$_j$] $\neq 0$, and again, the 
modified Tinsley equations are to be solved with yields for $Z>0$ and [X$_i$/X$_j$] $\neq 0$, ... . 

It is seen that via the SFH galaxy evolution and stellar evolution become intimately coupled. In principle, 
stellar evolutionary tracks and yields would be required not only for various metallicities (and He contents), 
but also for various abundance ratios. The SFH -- if short and burst-like or mild and $\sim const$ -- leads 
to different abundance ratios between elements with different nucleosynthetic origin, as e.g. [C/O] or 
[O, Mg, ..../Fe], where C comes from intermediate mass 
stars and Fe has important SNI contributions (cf. Sect.5), both
leading to a delayed production with respect to the SNII products O,
Mg, etc. However, no complete grid of stellar evolutionary tracks or yields for varying abundance ratios 
[Mg/Fe] or [${\rm \alpha/Fe}$] is 
available. 

Our chemically consistent ($=$ {\bf cc}) evolutionary synthesis models follow
the evolution of ISM abundances together with the spectrophotometric
properties of galaxies and account for the increasing initial
metallicity of successive generations of stars by using various sets
of input physics -- stellar evolutionary tracks, 
model atmosphere spectra, 
color and absorption index calibrations, yields, lifetimes, and remnant masses 
-- ranging in metallicity from ${\rm Z=10^{-4}}$ up to ${\rm Z=0.05}$. 

The two basic parameters of our evolutionary synthesis model are the IMF, which we take from Scalo, and 
SFHs, which we appropriately chose for different galaxy types. For ellipticals we use  
${\rm \Psi(t) \sim e^{-t/t_{\ast}}}$, for spiral types Sa ... Sc  ${\rm \Psi(t) \sim \frac{G}{M} (t)}$, 
and for Sd ${\rm \Psi(t) = }const.$ with characteristic timescales for SF ${\rm t_{\ast}}$ 
(for spirals defined via 
${\rm \int_0^{t_{\ast}} \Psi \cdot dt = 0.63 \cdot G\mid_{t=0}}$) ranging from 1 Gyr (E) 
to 2, 3, 10, and 16 Gyr for Sa, Sb, Sc, and Sd, respectively. 

Our chemically consistent models simultaneously describe 
{\bf the spectrophotometric evolution} in terms of 
spectra, luminosities, colors (UV -- IR), emission and absorption lines as a function of time or -- 
for any cosmological model as given by ${\rm H_0, \Omega_0, \Lambda_0,}$ and a redshift of galaxy 
formation ${\rm z_{form}}$ -- the redshift evolution of apparent magnitudes UBVRIJHK and 
colors, including 
evolutionary and cosmological corrections, as well as the attenuation by intervening HI, {\bf and 
the chemical evolution} in terms of ISM abundances of individual elements 
$^{12}$C, ..., $^{56}$Fe as a function of time or, again, of redshift. 

The SFHs have been chosen as to provide agreement, together with the IMF, of our model galaxies 
after a Hubble time with integrated colors, luminosities, absorption features (E/S0s), 
emission line strengths (spirals), typical for the respective galaxy types, as well as with 
template spectra (Kennicutt 1992, see M\"oller \etal 1999a for details), and characteristic 
HII region abundances. 

As compared to models using solar metallicity input physics only, our cc models use somewhat 
different SFHs, even for galaxies that by today come close to ${\rm Z_{\odot}}$. Clearly, 
differences between cc and ${\rm Z_{\odot}}$ models become increasingly important towards 
higher redshift for all galaxy models. 

We caution that our models are simple 1-zone descriptions without any
dynamics or spatial resolution, meant to describe global average
quantities like integrated spectra or colors, and absorption line 
strengths or HII region abundances around ${\rm \sim 1~R_e}$. 

\section{Chemically Consistent Photometric Evolution}
Having only available a very incomplete grid of stellar evolutionary 
tracks and color calibrations, 
Arimoto \& Yoshii 1986 were the first to attempt a cc approach to photometric evolution. 
Einsel \etal 1995 compiled more complete data sets for 5 metallicities to describe
the photometric evolution in terms of colors and absorption line
indices on the basis of stellar tracks from the Geneva, color
calibrations from the Yale, and index calibrations from the Lick groups, respectively. 
In M\"oller \etal 1997 we present cc spectrophotometric evolution
models based on most recent input physics. We discuss the comparison 
between the (mass-weighted) ISM
metallicities  for various galaxy types and the luminosity-weighted
metallicities of the stellar population seen in different wavelength
bands. At late stages, stars in models with $const.$ SF (Sd) show a
stellar metallicity distribution strongly peaked at ${\rm \frac{1}{2}
Z_{\odot}}$ at all wavelengths and close to the ISM
metallicity. Elliptical models, on the other hand, show broad stellar
metallicity distributions extending from ${\rm Z=10^{-4}~to~Z=0.05}$
in all bands with the maximum luminosity contribution in U coming from stars
with ${\rm < \frac{1}{2}Z_{\odot}}$, and in K from stars with ${\rm
\sim Z_{\odot}}$, as shown in Fig.1. The largest difference of as much as a factor 2 is obtained 
in our Sb model between the luminosity-weighted stellar metallicity (e.g. in V) and 
the ISM metallicity. The time evolution of all these metallicities for
different galaxy types are presented in M\"oller \etal 1997 together
with the evolution of colors and absorption line indices. In Kurth
\etal 1999 we present single burst single metallicity models using
stellar tracks from the Padova group and compare the time evolution of
colors and absorption indices to globular cluster observations. We give theoretical color and 
index calibrations in terms of [Fe/H] and investigate how they evolve with age.  

\begin{figure}
\vspace{-6pt}
\centerline{\psfig{figure=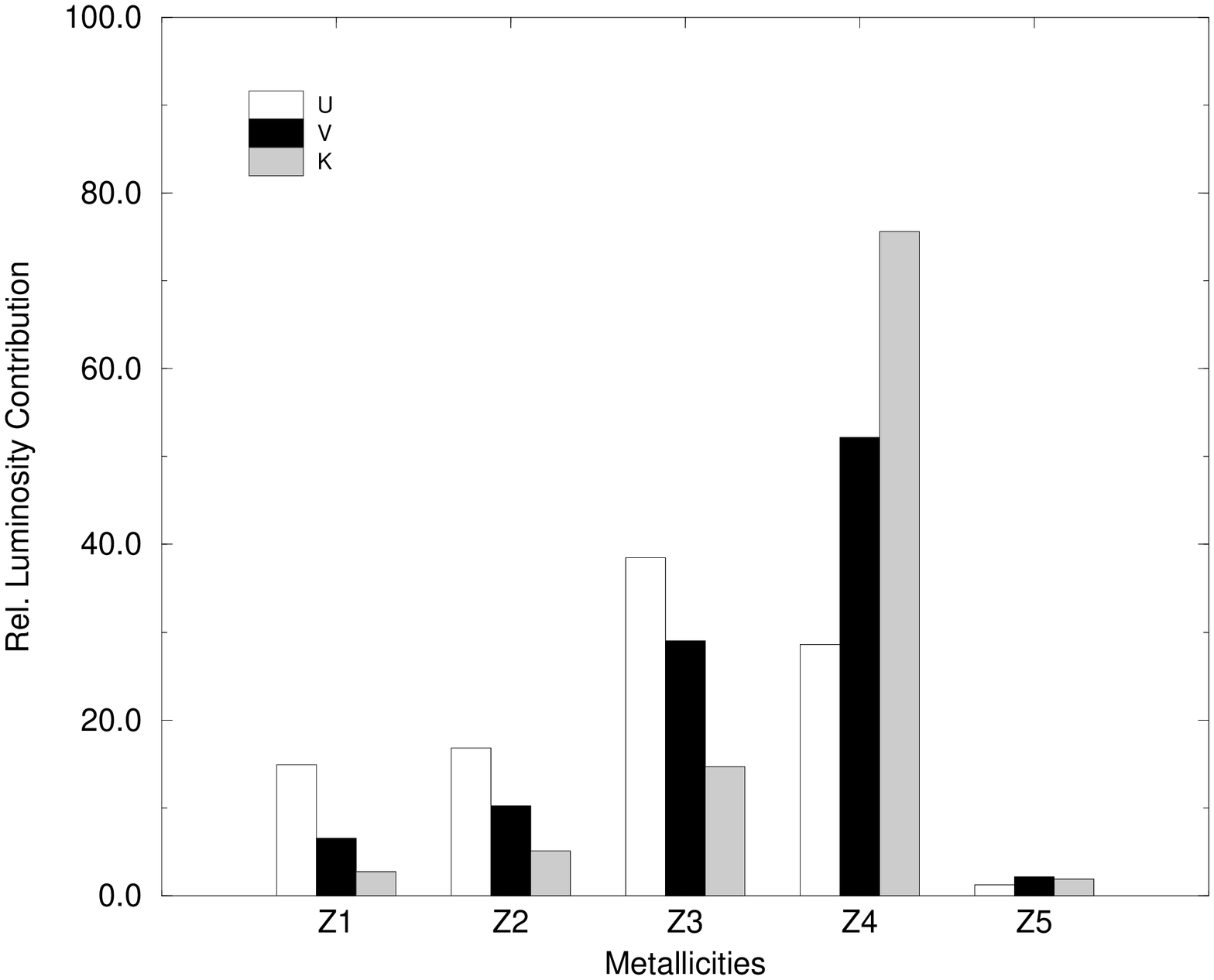,height=6cm,angle=270}\psfig{figure=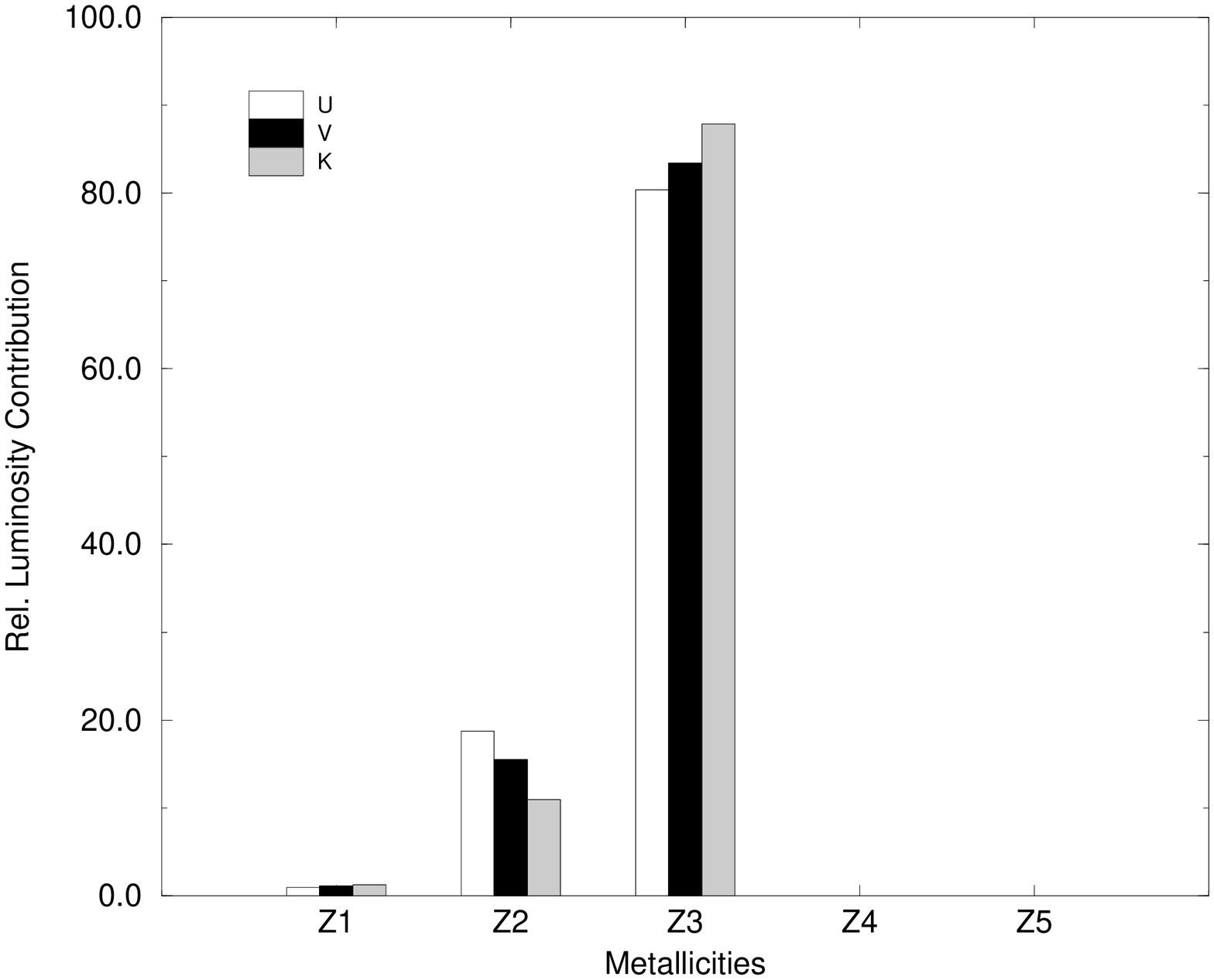,height=6cm,angle=270}}
\vspace{-5pt}
\caption{Metallicity distributions of stars in E (Fig.1a) and Sd
galaxies (Fig.1b) as
seen in U, V, and K bands. Metallicities are ${\rm Z1=0.0004,~Z2=0.004,~Z3=0.008,~Z4=0.02,~Z5=0.05}$.}
\vspace{-6pt}
\end{figure}

Stellar abundance ratios, like [Mg/Fe]$_{\ast}$, as derived from absorption line ratios, 
depend on the initial 
abundance ratio ${\rm [Mg/Fe]_{\ast}}$ of the star, on 
modifications through nucleosynthesis and mixing during 
the life of the star, and on physical parameters in the stellar atmosphere. The Fe index, e.g., 
is as sensitive to Fe as to global metallicity Z, which to 50\% is made up by oxygen from SNII 
(Tripicco \& Bell 1995). 
Abundance ratios in galaxies additionally depend on the age and metallicity distributions of the stars 
and, hence, on the SFH, the IMF, any possible pre-enrichment (from Pop 3, halo, ...), on metal-poor 
infall and/or metal selective outflow, which determine the ISM abundance ratio 
${\rm [Mg/Fe]_{ISM}}$ before the birth of the stars. In particular do solar abundance ratios, taken 
for reference, reflect the local SFH, IMF, and all pre-enrichment \& dilution effects of the local ISM. 
Looking at the metallicity dependent SNII -- yields (cf. Sect.5) we notice that, integrated over 
the IMF, the ejected mass ratio ${\rm
M(Mg)/M(Fe)}$ increases by large factor when going down in metallicity from 
${\rm Z_{\odot}}$ to ${\rm 10^{-3}\cdot
Z_{\odot}}$. A lower metallicity limit for SNIa, as discussed by
Kobayashi \etal 1998 would further increase ${\rm [Mg/Fe]_{ISM}}$ at
early stages. Therefore, all conclusions drawn for galaxies from a comparison of 
observed ${\rm [Mg/Fe]_{\ast} > 0}$ with ${\rm Z_{\odot}}$ models -- concerning 
a top-heavy IMF, a shorter SF timescale in massive Es, a higher SFE,
etc. --  
should be taken with extreme caution (cf. Fritze - v. Alvensleben 1998).

\section{Chemically Consistent Spectro-Cosmological Evolution}
Using sets of model atmosphere spectra covering all spectral types,
luminosity classes, and our 5 metallicities (Lejeune \etal 1997, 1998)
we describe the cc spectral evolution of our model galaxies. The
agreement of our model spectra after a Hubble time with nearby
templates is shown by M\"oller \etal 1997 and 1999a. With any kind of
cosmological model we 
calculate evolutionary and cosmological corrections, apparent magnitudes U, ..., K, and colors 
as a function of redshift z. Attenuation by intervening hydrogen (Madau 1995) is included, dust 
extinction depending on metallicity and gas content is currently being added in collaboration 
with D. Calzetti (cf. M\"oller \etal 1999b). 

\begin{figure}
\vspace{-6pt}
\centerline{\psfig{figure=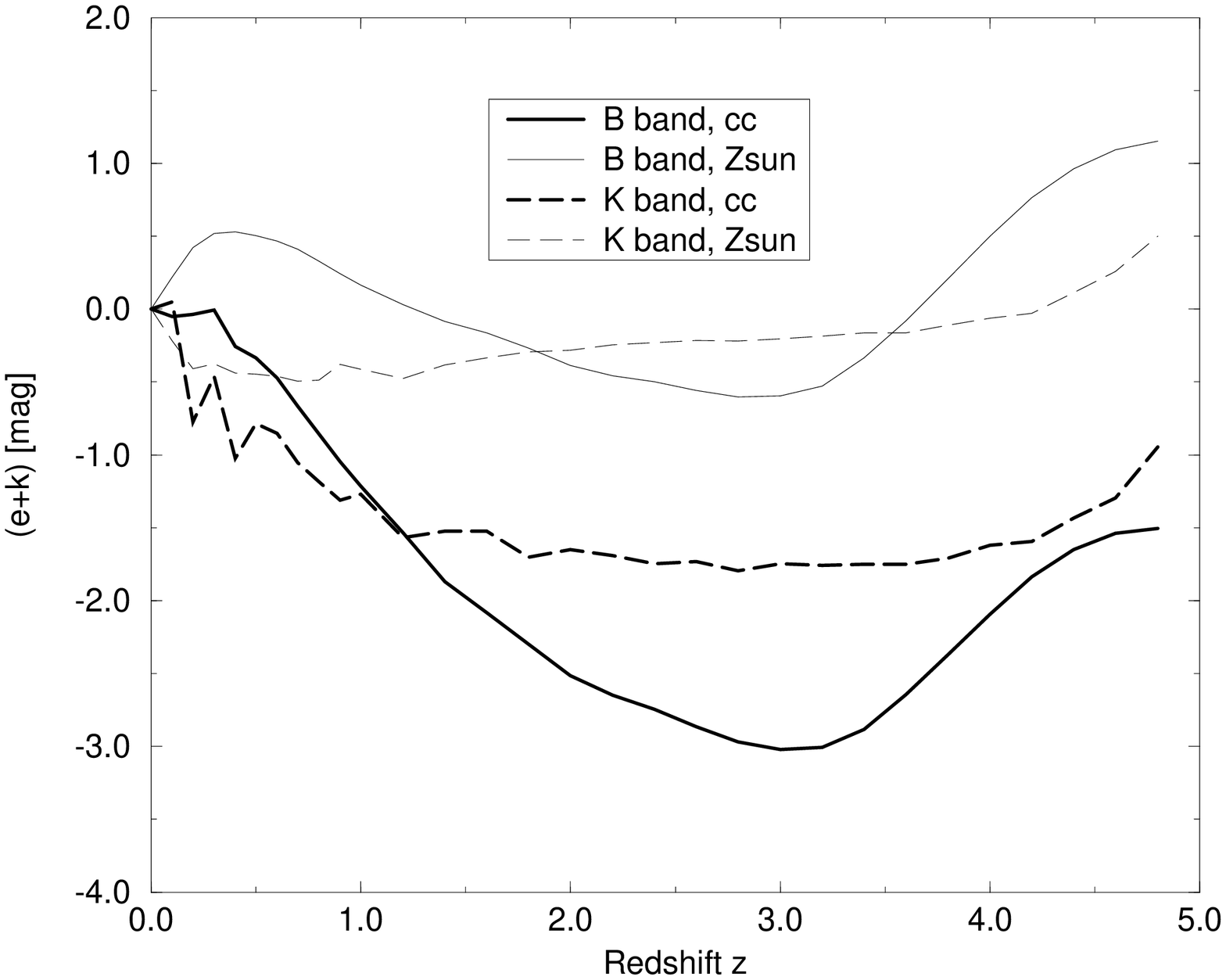,height=6cm,angle=270}\psfig{figure=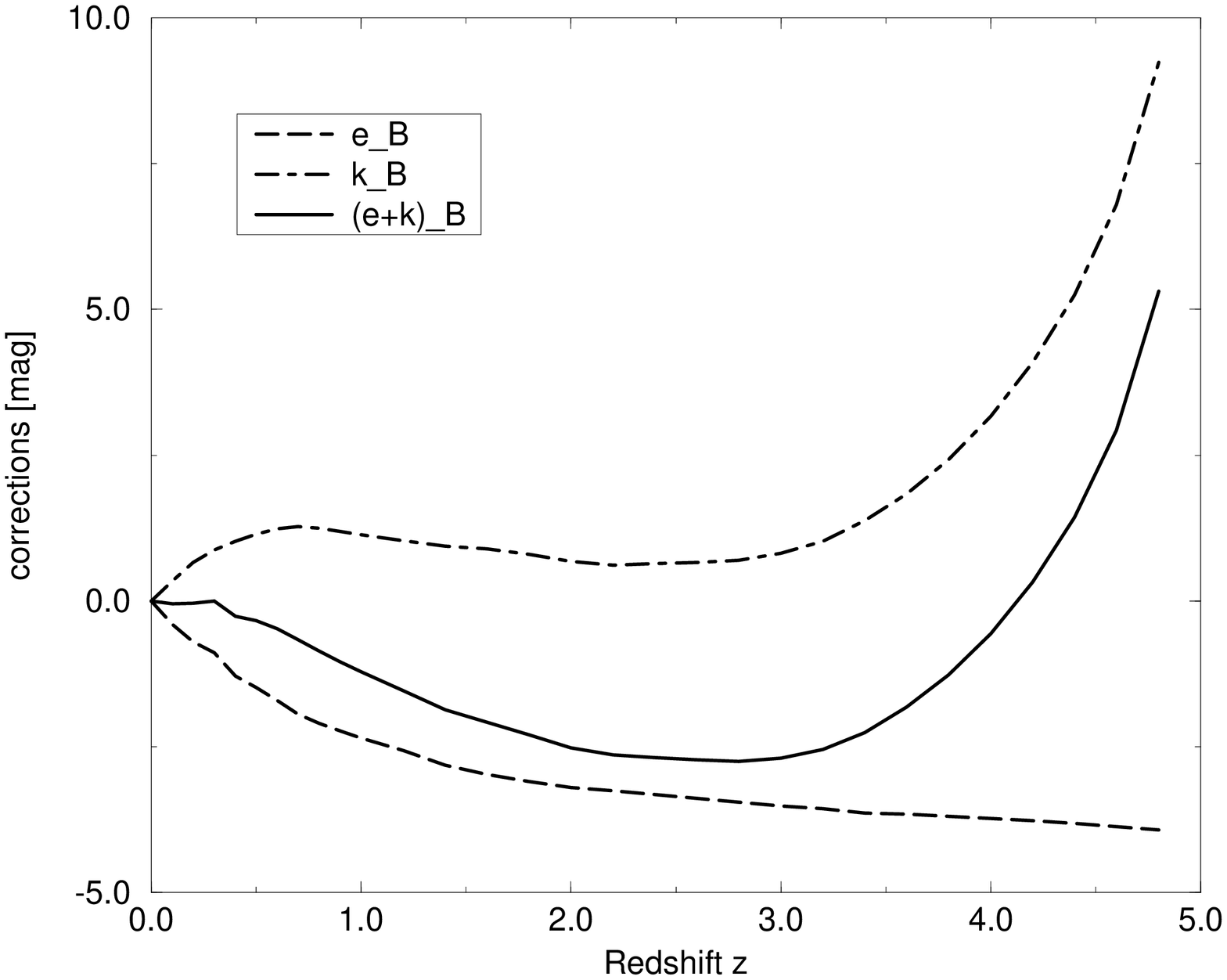,height=6cm,angle=270}} 
\vspace{-5pt}
\caption{Comparison of ${\rm (e+k)}$ corrections without attenuation in B and K for Sd galaxies in cc
and ${\rm Z_{\odot}}$ models (Fig.2a). e-, k-, and ${\rm (e+k)}$ - corrections for Sd
galaxies in B from cc models with attenuation included (Fig.2b). ${\rm
((H_o,\Omega_o,\Lambda_o)=(50,1,0),~ z_{form}=5)}$.}
\vspace{-6pt}
\end{figure}

For the Sd galaxy model, Fig.2a shows the cc ${\rm (e+k)}$ corrections in B
and K as compared to those for the ${\rm Z_{\odot}}$ model. 
In B, cc models give ${\rm (e+k)}$ corrections that make Sd
galaxies brighter by $\gta 1$ mag at ${\rm z \sim 0.7}$
and by $\gta 2$ mag for ${\rm z \gta 2}$ 
as compared to ${\rm Z_{\odot}}$ models. In the K-band
as well, differences are quite significant with Sd galaxies being
brighter in cc models by $\gta 1$ mag at ${\rm  z \gta 1}$. 
Moreover, the successive building-up of the stellar metallicity
distributions tends to increase the evolutionary effects as compared
to ${\rm Z_{\odot}}$ models. In addition to the cosmological dimming
(${\rm k_{UBVRI}>0}$) the Sd model is brightened by the evolutionary
correction by $\gta 2$ mag at ${\rm z \gta 0.7}$ and by $\sim 3.5$ mag at
${\rm z \sim 3}$ in B (cf. Fig.2b), while in K, Sd models still have ${\rm
(e+k) \lta -1.5}$ mag at all ${\rm z \sim 0.7 - 3}$. 
The cc spectro-cosmological models and the comparison with high redshift galaxy data 
will be presented in detail in M\"oller \etal ({\sl in prep.}).

\section{Chemically Consistent Chemo-Cosmological Evolution}
Timmes \etal 1995 and Portinari \etal 1998 were the first to use stellar yields for a range of 
metallicities in models for the chemical evolution of the Milky Way and the solar 
neighbourhood, respectively. 
For our ``spectroscopically successful'' models for various galaxy types we now use element 
yields for different metallicities for successive generations of stars to describe the time 
evolution of a series of individual elements from $^{12}$C through $^{56}$Fe. For any 
cosmological model the time evolution directly transforms into a redshift evolution. 
SNII yields for massive stars (${\rm >8~M_{\odot}}$) are from Woosley \& Weaver 1995, 
yields for intermediate mass stars from van den Hoek \& Groenewegen 1997. SNIa contributions 
to Fe, C, ..., are included for the carbon deflagration white dwarf binary scenario as outlined 
by Matteucci \& Greggio 1986. SNIa yields are only available for ${\rm Z_{\odot}}$ (Nomoto 
\etal 1997, model W7), however, no important metallicity dependence is expected for SNIa yields 
except for a possible lower metallicity limit to the explosion (Kobayashi \etal 1998). Mass loss 
from stellar winds as e.g. given by Portinari \etal 1998 is not included yet. We recall that 
metallicity dependent stellar yields are only available for solar abundance ratios and we 
have no idea, if and in how far non-solar abundance rations would influence the stellar 
shell structure and, hence, the yields. Moreover, stellar yields depend on 
${\rm \frac{\Delta Y}{\Delta Z}}$, explosion energies, remnant masses, etc. 
and the metallicity dependence of these factors is still poorly understood. No clear trends 
are seen, neither in the output of element ${\rm X_i}$ from stars of given mass as a function 
of metallicity, nor, at const. metallicity, as a function of stellar mass. 

We stress that for a given SF history and IMF our models yield absolute abundances that do not 
require any scaling or normalisation. 

We compared the redshift evolution of C and Mg abundances with CIV-
and MgII-QSO absorber statistics (Fritze - v. Alvensleben \etal 1989,
1991 using yields for ${\rm Z_{\odot}}$ stars only). CIV- and
MgII-absorption is caused by the moderate column density gas in
extended galaxy halos (${\rm 17.5 \leq 
log~N(HI)~[cm^{-2}] \leq 20}$). We argued that the cross section for CIV- and
MgII-absorption should scale with the abundance of the respective
elements and that, hence, the redshift evolution of the comoving
number density of absorbers should trace the abundance evolution. The first direct abundance 
determinations (Stengler -
Larrea 1995) {\sl a posteriori} justified our assumption. We found
good agreement of the observations with a standard model for halo SF
(${\rm t_{\ast} \sim 1~Gyr}$), derived constraints for ${\rm t_{\ast}}$, the
IMF, and the cosmological parameters, and predicted a low number of CIV systems at low redshift, 
which was impressively confirmed by HST key project data (Bahcall \etal 1993). 

Our cc chemo-cosmological models are compared with observed abundances
in damped Ly$\alpha$ absorbers (=DLAs) in Lindner \etal 1999. DLAs
show radiation damped Ly$\alpha$ lines due to high column density gas
(${\rm log~N(HI)~[cm^{-2}] \geq 20.3}$) and a large number of associated
low ionisation
lines of C, N, O, Al, Si, S, Cr, Mn, Fe, Ni, Zn, ... High resolution observations (KECK and WHT) 
that fully resolve the complex velocity structure in the lines allow to derive precise element 
abundances in a large number of DLAs over the redshift range 0 ... $\geq 4$ (Boiss\'e \etal 1998, 
Lu \etal 1993, 1996, Pettini \etal 1994, 1999, 
Prochaska \& Wolfe 1997, ...). Based on similarities of their HI
column densities with those of 
local spiral disks, of their comoving gas densities at high z with
(gas $+$ star)-densities in local galaxies, and based on line
asymmetries indicative of rotation,  damped Ly$\alpha$ absorption is
thought to arise in (proto-)galactic disks along the line of sight to
a distant QSO (e.g. Wolfe 1995). Alternatively, Matteucci \etal 1997
propose starbursting dwarf galaxies on the basis of [N/O] ratios,
while Jimenez \etal 1999 propose LSB galaxies, and Haehnelt \etal 1998
subgalactic fragments to explain DLA galaxies at low and high redshift, respectively.  

After referring all observed DLA abundances to one homogeneous set of
oscillator strengths and solar reference values, we compare with our
spiral galaxy models Sa, ..., Sdearlier shown to 
agree with characteristic HII region abundances at ${\rm z=0}$, and
with spectrophotometric properties as observed to ${\rm z
\gta 1}$. As can be seen in Fig.3 on the example of Zn, our Sa and Sd
models bracket the redshift evolution of DLA abundances from ${\rm z
\geq 4.4}$ to ${\rm z \sim 0.4}$. Similar agreement is found for all
8 elements with a reasonable number of DLA data available.
Since Phillipps \& Edmunds 1996 and Edmunds \& Phillipps 1997 have
shown that the probability for an arbitrary QSO sightline to cut
through an intervening gas disk and produce DLA absorption is highest
around ${\rm 1~R_e}$, our 
models bridge the gap from high-z DLAs  to 
nearby spiral HII region abundances. We conclude that {\bf from the
point of view of abundance evolution, DLA galaxies may well be the
progenitors of normal spirals Sa -- Sd}, allthough we cannot exclude
that some starbursting dwarf or LSB galaxies may also
be among the DLA galaxy sample.

\begin{figure}
\vspace{-6pt}
\centerline{\psfig{figure=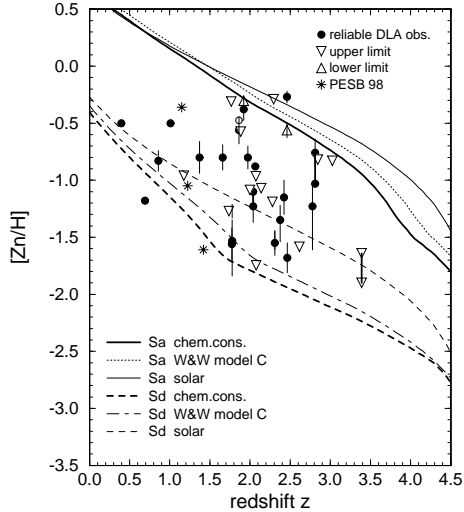,height=7cm}} 
\vspace{-5pt}
\caption{Redshift evolution of [Zn/H]. 
Heavy lines are for our chemically consistent models, thin lines are for models 
using solar metallicity input physics only. 
}
\vspace{-6pt}
\end{figure}

The influence of the metallicity dependent yields is seen from the
comparison with ${\rm Z_{\odot}}$ models in Fig.3 and varies from
element to element. Whenever a significant difference is seen, the cc
models give a good representation of the data while the ${\rm
Z_{\odot}}$ models lose data points above the Sa or below the Sd
curves. We also show models with SNII yields calculated under the
assumption of higher explosion energies (model C of Woosley \& Weaver
95), which does not make much difference. Curves for Sb and Sc models run between those for 
Sa and Sd and are omitted for clarity. 

Comparison of cc chemo-cosmological models with observed DLA abundances further shows that the 
weak redshift evolution of DLA abundances is a natural result of the 
long SF timescales for disks galaxies, and the 
range of SF timescales 
${\rm t_{\ast}}$ for spirals from Sa through Sd fully explains the abundance 
scatter among DLAs at any redshift. 

Somewhat surprisingly, abundances of elements which locally are known
to strongly deplete onto dust grains (like Fe or Cr) are as well
described by our models as are non-refractory elements like Zn. We
prefer, however, not 
to draw conclusions about the importance of
dust in DLAs in view of the uncertainties in the stellar yields and the simplicity of our 
closed-box 1-zone model. 

Fig.3 also shows that while at high redshift all spiral types seem to
give rise to DLA absorption, no more data points at ${\rm z \lta 1.5}$
reach close to our early type spiral models. {\bf At low redshift, the gas poor
early type spirals seem to drop out of DLA samples}. While a deficiency of high N(HI) systems 
at low z has been noted before (Lanzetta \etal 1997), and attributed to their high metallicity 
and dust content (Steidel \etal 1997), our models indicate an additional reason: as the global 
gas content drops, the probability for a QSO sightline to cut through a high N(HI) part of the 
galaxy decreases, i.e. the cross section for {\bf damped} Ly$\alpha$ absorption gets reduced. 
If this were confirmed by further low-z DLA data, it would have serious implications as to the 
posibility to optically identify DLA galaxies. Locally, on average, Sd galaxies are fainter by 
$\sim 2$ mag in B than Sa's and our cc spectro-cosmological models predict that the low-z DLA 
galaxies should be about as faint in B, \R, and K as the brightest members of the high-z population: 
${\rm B \sim 25}$, \R $~\sim 24.5$, K $\sim 22$ mag. Luminosities of
the few optically identified DLA galaxies (and candidates) to date are
in good agreement with our predictions (cf. Fritze - v. Alvensleben
\etal 1999a, b, c).

Optically identified DLA absorbers with information about ISM
abundances from the metal absorption lines {\bf and}
spectrophotometric properties of the stellar population allow to much
better constrain the model parameters than either aspect (ISM or
stars) alone (cf. Lindner \etal 1996). Since DLA galaxies are within the reach of 10m-class 
telescopes up to redshifts z $> 3$ and trace the normal galaxy population to these high 
redshifts without any bias as to high luminosity, radio power, or the like, they can 
give powerful constraints on the evolutionary histories and ages, and, ultimately, even 
on the cosmological parameters. Accurate abundance data in very low metallicity DLAs may 
provide valuable clues for the nucleosynthesis at low metallicity.

\section{Conclusions and Outlook}
I pointed out the importance of a chemically consistent modelling of the spectrophotometric 
and chemical evolution of galaxies that takes into account the evolving metallicity 
distribution of the stellar population. The comparison with models using solar metallicity 
input physics only showed substantial differences due to important contributions of subsolar 
metallicity stars both to the yields of various elements and to the spectrophotometric 
properties for all galaxy types, in particular when going to high redshift. I stressed the 
need for reliable low metallicity input physics (stellar tracks, mass loss, lifetimes, yields, 
remnant masses, spectra, and absorption index calibrations) and cautioned the use of solar 
metallicity models to derive conclusions from non-solar absorption line ratios in the clearly 
composite stellar populations of E/S0 galaxies.
Observations of the stellar metallicity distributions in nearby galaxies will be necessary 
before we can reasonably relax the crude simplification of closed box models. With the same 
number of parameters (IMF and SF history) as standard chemical or spectrophotometric models, 
our unified chemical, spectrophotometric and cosmological model provides a powerful tool to 
constrain the parameters and ages of galaxies for which both kind of information is available.

\acknowledgments
I gratefully acknowledge partial financial support from the 
Organisers and from the Deutsche Forschungsgemeinschaft (Fr 916/7-1).

\end{document}